\documentclass[preprint]{aastex}

\input{psfig.sty}
\def\arcsec {$^{\prime \prime}$}
\def\arcmin {$^{\prime}$}

\def\cmmb   {~cm$^{-2}$}

\def\degr   {$^{\circ}$}
\def\etal   {{\it et~al.\/}}

\def\kms    {~km~s$^{-1}$}
\def\mo     {{$M_{\odot}$}}

\def\spitz {{\it Spitzer}}
\def\xiiico  {{$^{13}{\rm CO}$}}

\begin{document}

\title{Discovery of a New Low-Latitude Milky Way Globular Cluster
using GLIMPSE}

\author{ Henry A. Kobulnicky, \altaffilmark{1} A.~J.  Monson, \altaffilmark{1}
B.~A Buckalew, \altaffilmark{1}
J.~M. Darnel, \altaffilmark{1}
B. Uzpen, \altaffilmark{1}
M. R. Meade,\altaffilmark{2} 
B. L. Babler,\altaffilmark{2}
R. Indebetouw,\altaffilmark{2}
B. A. Whitney,\altaffilmark{3} 
C. Watson,\altaffilmark{2}
E. Churchwell\altaffilmark{2}
M. G. Wolfire,\altaffilmark{4}
M. J. Wolff,\altaffilmark{3}
D. P. Clemens,\altaffilmark{5} 
R. Shah,\altaffilmark{5} 
T. M. Bania,\altaffilmark{5} 
R. A. Benjamin,\altaffilmark{6}
M. Cohen,\altaffilmark{7} 
J. M. Dickey,\altaffilmark{8}
J. M. Jackson,\altaffilmark{5} 
A. P. Marston,\altaffilmark{9}
J. S. Mathis,\altaffilmark{4} 
E. P. Mercer,\altaffilmark{6}
J. R. Stauffer,\altaffilmark{10} 
S. R. Stolovy,\altaffilmark{10}
J.~P. Norris,\altaffilmark{11} 
A. Kutyrev,\altaffilmark{12} 
R. Canterna,\altaffilmark{1} 
M.~J. Pierce\altaffilmark{1} 
}

\altaffiltext{1}{University of Wyoming, Dept. of Physics \& Astronomy, PO
Box 3905, Laramie, WY 82072}

\altaffiltext{2}{University of Wisconsin-Madison, Dept. of Astronomy,
475 N. Charter St., Madison, WI 53706}

\altaffiltext{3}{Space Science Institute, 4750 Walnut St. Suite 205,
Boulder, CO 80301}

\altaffiltext{4}{University of Maryland, Dept. of Astronomy, College
Park, MD 20742-2421}

\altaffiltext{5}{Boston University, Institute for Astrophysical
Research, 725 Commonwealth Ave., Boston, MA 02215}

\altaffiltext{6}{University of Wisconsin-Whitewater, Physics Dept.,
800 W. Main St., Whitewater, WI 53190}

\altaffiltext{7}{University of California-Berkeley, Radio Astronomy
Lab, 601 Campbell Hall, Berkeley, CA 94720}

\altaffiltext{8}{University of Minnesota, Dept. of Astronomy, 116 Church
St., SE, Minneapolis, MN 55455}

\altaffiltext{9}{ESTEC/SCI-SA,Postbus 299,2200 AG Noordwijk,The
Netherlands}

\altaffiltext{10}{Caltech, Spitzer Science Center, MS 314-6, Pasadena,
CA 91125}

\altaffiltext{11}{NASA Goddard Space Flight Center, LHEA, Code 661, Greenbelt, MD 20771}
\altaffiltext{12}{NASA Goddard Space Flight Center, SSAI, Code 685, Greenbelt, MD 20771}


\author{Accepted for Publication in {\it The Astronomical Journal}}

\vskip 1.cm

\begin{abstract}
\spitz\ Space Telescope imaging from the Galactic Legacy
Infrared Mid-Plane Survey Extraordinaire (GLIMPSE) reveals a previously
unidentified low-latitude rich star cluster near $l$=31.3\degr,
$b$=-0.1.  Near-infrared JHK' photometry from the Wyoming Infrared
Observatory indicates an extinction of $A_V\simeq15\pm3$ mag for
cluster members.  Analysis of \xiiico\ features along the same
sightline suggests a probable kinematic distance of 3.1 -- 5.2 kpc.
The new cluster has an angular diameter of $\sim1-2$ pc, a total
magnitude corrected for extinction of $m_{K_0}=2.1$, and a luminosity of
$M_K\simeq-10.3$ at 3.1 kpc. In contrast to young massive Galactic
clusters with ages $<$100 Myr, the new cluster has no significant
radio emission.  Comparison to theoretical K-band luminosity functions
indicates an age of at least several Gyr and a mass of at least $10^5$
\mo. Unlike known old open clusters, this new cluster lies in the
inner Galaxy at $R_{GC}\simeq6.1$ kpc.  We designate this object
``GLIMPSE-C01'' and present evidence that it is a Milky Way globular cluster
passing through the Galactic disk.  We also identify a region of star
formation and fan-shaped outflows from young stellar objects in the
same field as the cluster.  The cluster's passage through the Galactic
molecular layer may have triggered this star formation activity.
\end{abstract}


\section{Introduction}

Infrared surveys of the Galactic Plane, such as 2MASS, have led to the
discovery of several hundred new star clusters throughout the Milky
Way disk (e.g., Borissova \etal\ 2003; Bica \etal\ 2003; Dutra
\etal\ 2003; Hurt \etal\ 2000).  Typically, clusters are identified by
locating peaks in the surface density of stars using point source
catalogs.  The majority of these clusters were previously undetectable
in optical surveys due to high extinction at low latitudes.
Near-infrared cluster searches, however, are still incomplete in
regions of highest extinction where $A_K$ may reach several
magnitudes.

The Galactic Legacy Mid-Plane Survey Extraordinaire (GLIMPSE) is
mapping the Galactic Plane in four infrared array camera (IRAC; Fazio
\etal\ 2004) bands at 3.6 $\mu$m, 4.5  $\mu$m, 5.8  $\mu$m, and 8.0 
$\mu$m from $|l|=10^\circ-65^\circ$ and $|b|<1^\circ$ 
(Benjamin \etal\ 2003).  One of the primary science drivers for the
survey is to enable a complete census of star formation regions and
stellar populations throughout the inner Galaxy, unhindered by
extinction, at the same arcsecond angular resolutions as optical and
near-IR surveys.  The survey will also enable discovery of previously
uncatalogued stellar objects on the far side of the Galaxy or in
highly obscured regions.

In this paper we report the serendipitous discovery of a rich star
cluster in the first Galactic quadrant near $l=31$, $b=-0.1$.  The
cluster may be a new member of the collection of $\sim150$ (Harris
1996) Milky Way globular clusters.  We will refer to this object as
GLIMPSE-C01.

\section{Data}

\subsection{GLIMPSE Imaging and Photometry}

The segment of the the Galactic Plane from $l=25$\degr\ to
$l=40$\degr\ was observed by \spitz\ with the IRAC camera on 2004
April 21 as part of the GLIMPSE program.  The total exposure time at each
IRAC band is 2.4 seconds and the instrumental resolution ranges
from 1.6\arcsec\ FWHM at 3.6 $\mu$m to 1.9\arcsec\ FWHM at 8.0 $\mu$m.
Mosaiced images of the 5\arcmin$\times$5\arcmin\ IRAC frames were
constructed by the GLIMPSE team using MONTAGE.\footnote{Montage 
software is funded by the National Aeronautics and Space Administration's 
Earth Science Technology Office.}
A member of the GLIMPSE team (A.~J.~M.) identified the cluster 
during visual examination of
the images.   The only previous identification of this object
was by Simpson \& Cotera (2004) using 2MASS near-infrared images
to follow up on ASCA X-ray detections in the Galactic Plane. 
Inspection of initial 4.5 $\mu$m mosaiced images over the
regions $l=25-40$\degr and $l=306-337$\degr shows that {\it no other
similarly rich bright clusters are found in the 92 square degrees of the
Plane surveyed to date.}  The center of GLIMPSE-C01 is at
$l=31.30$\degr, $b=-0.10$\degr or RA(J2000)=18h48m49.7s,
DEC(J2000)~=~-01d29m50s.  For further details on GLIMPSE imaging and
photometry see Mercer \etal\ (2004), Churchwell \etal\ (2004), Whitney
\etal\ (2004), or Indebetouw \etal\ (2004).

Figure~\ref{IRAC} shows logarithmic greyscale representations of
images in each of the 4 IRAC bands. The IRAC1 and IRAC2 bands (upper
panels) are dominated by stellar photospheric emission from normal
main sequence stars.  The cluster is the most prominent feature in the
field.  It subtends over 2 arcminutes on the sky and is dominated by 3
bright pointlike sources which we show later are multiple blended
stars.  We note that the brightest probable members are located near
the cluster center, suggesting an age old enough for significant
dynamical evolution to have occurred.  Stars become less prominent in
IRAC bands 3 \& 4 which are increasingly dominated by emission from
known polycyclic aromatic hydrocarbon (PAH) bands.  The lower panels
of Figure~\ref{IRAC} reveal a bright swath (hereafter the ``plume'') of
diffuse emission extending $\sim1$\arcmin\ toward the Galactic south
and toward higher longitudes.  These panels also show a region
(hereafter the ``ribbon'') of lower surface brightness diffuse
emission running several arcminutes from the cluster toward the
Galactic north and toward lower longitudes.
 
Figure~\ref{3color} is a 3-color image of GLIMPSE-C01 and the
surrounding field constructed from the IRAC1 3.6 $\mu$m image (blue),
the IRAC3 5.8 $\mu$m image (green) and the IRAC4 8.0 $\mu$m (red).
Contours indicate the 1420 GHz radio continuum emission in the NRAO
VLA Sky Survey (Condon \etal\ 1998) archival image with contours at 2,
3, 4, 5, and 10 times the $1\sigma$ RMS noise of 1.3 mJy/beam.  The
cluster is coincident with a $3\sigma$ radio detection at a level of
4.5 mJy per 45\arcsec\ synthesized beam.  Reduction of VLA
B-configuration 1400 MHz archival data from program code AC629
(10\arcsec\ synthesized beam) shows no point source at this location
to a limit of 1.3 mJy, indicating that this radio emission in the NVSS is
probably diffuse on scales larger than 10\arcsec.  Several stronger
(presumably unrelated) radio sources are visible to the Galactic S and
toward lower longitudes from the cluster.

We note in Figure~\ref{3color} the presence of several Y-shaped features
2\arcmin\ south (Galactic) of the cluster.  These objects appear similar
to outflow cones from young stellar objects, and probably indicate
a region of star formation.  In Section 2.5 below we show that this region
is coincident with a peak in the \xiiico\ surface brightness at a velocity
of 46 \kms, suggesting a kinematic distance of 3.1 kpc.  It is not clear 
whether this star formation region is affiliated with the cluster or
is a foreground or background object.

\subsection{WIRO Near-IR Imaging}

Near-infrared imaging of the GLIMPSE-C01 was obtained 2004 July 31
using the 256$^2$ InSb Goddard Infrared Camera (GIRcam) on the 2.3 m
Wyoming Infrared Observatory (WIRO) telescope.  GIRcam has a pixel
scale of 0.46\arcsec\ pixel$^{-1}$ at the Cassegrain focus.  Seeing
was 1.1-1.2\arcsec. 
Images were obtained in the J, H, and K' filters with total exposure
times of 120 s, 200 s, and 320 s respectively, broken into multiple
``dithered'' exposures.  Background images were obtained every 60 s on
adjacent regions of sky.  Data reduction followed standard procedures.
Sky background exposures before and after each sequence were averaged
and subtracted from each on-source image.  Flat field images in each
filter were constructed from a median of at least 20 on-sky exposures
obtained throughout the night.  Cluster frames were registered and
combined to produce final images.  Conditions were photometric, and 3
standard stars from the list of Hawarden
\etal\ (2003) covering a range of J-K color
were observed for flux calibration.  A collimation problem with the
secondary mirror produced stellar images with a narrow core (FWHM of
2.5 pixels or 1.15\arcsec) but broad, asymmetric wings.  Eighty three
percent of the power from a point source is contained in the core
within a 4 pixel radius. 
 DAOphot PSF fitting photometry was performed
on the images to obtain JHK' photometry down to limiting magnitudes of
17.2, 15.5, and 14.5 in the J,H,K' bands, respectively.  In each
filter 4-5 isolated stars were chosen to create a PSF profile that was
applied to the final images. Three deconvolution iterations were
performed on the cluster to obtain magnitudes for 313 stars.  
The photometry is complete to a K-band magnitude of $\sim$12.5.
We transformed the JHK' magnitudes from the UKIRT photometric system to the 2MASS
system using the relations of Carpenter (2001).

\begin{figure}
    \plotone{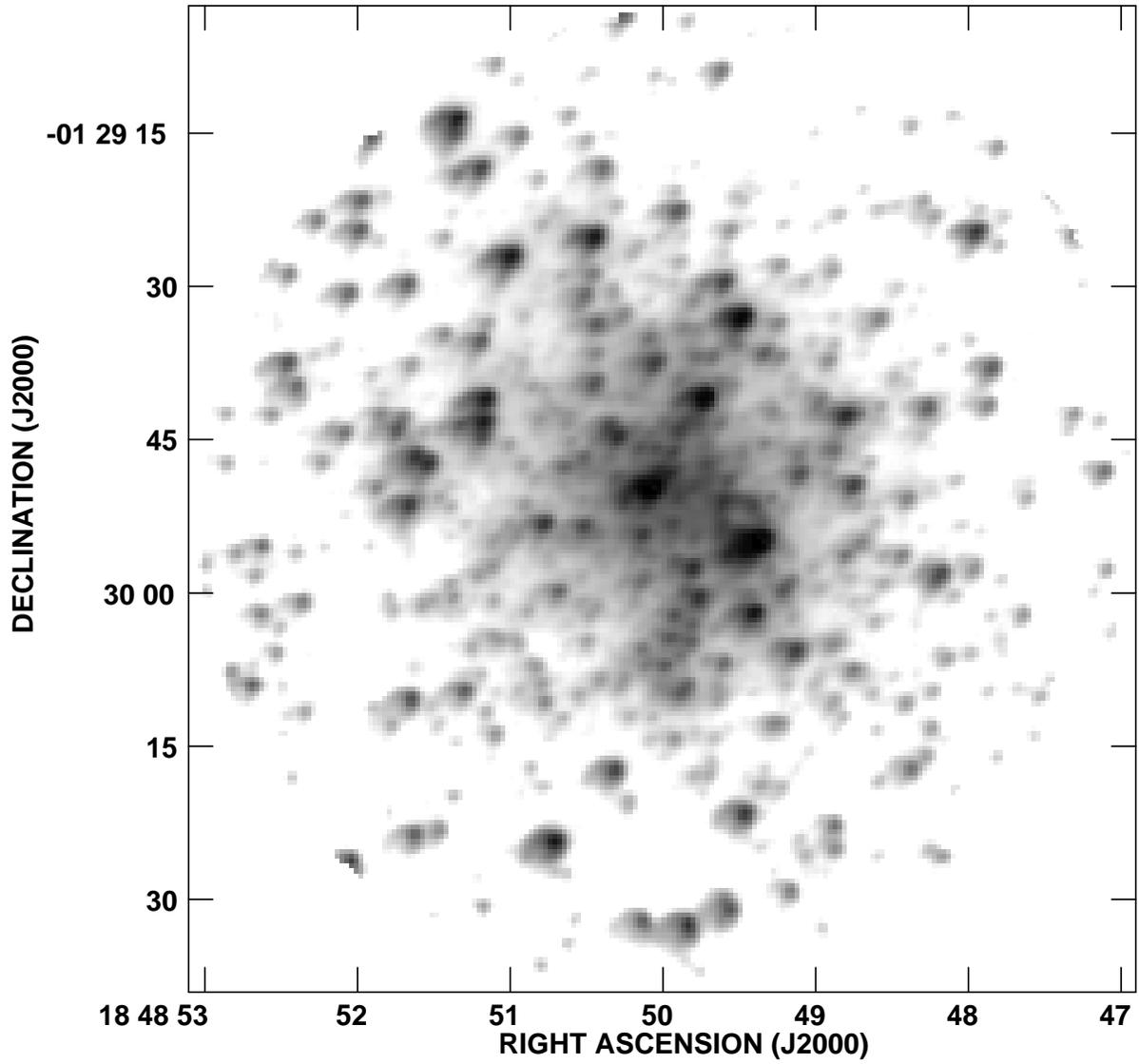}
    \caption{A logarithmic greyscale representation showing the WIRO H-band image
of GLIMPSE-C01.   \label{wiro1} }
\end{figure}

Figure~\ref{wiro1} shows a WIRO H-band image of
the cluster.  Although the cluster is readily visible in 2MASS J,H,K' images,
it has not been previously identified in the literature.  The WIRO 
data analysed here have superior sensitivity and angular resolution
than the 2MASS archival images.

\subsection{Integrated Luminosity}

We measured the cluster's total flux by performing aperture photometry
on the 2MASS JHK' images and our \spitz\ IRAC 1-4 images using a
90\arcsec\ radius aperture and an annular background region from
90\arcsec\ to 110\arcsec\ from the cluster.  Table~1 lists the
integrated magnitudes and fluxes at each band.  Note that the flux in
IRAC bands 3 \& 4 contains a considerable contribution from diffuse
PAH emission (seen in Figure~\ref{IRAC}) which is not present at other
wavelengths.

\subsection{Far Infrared IRAS }

An examination of archival images from the IRAS mission shows the new
cluster to be located near the periphery of an extended region of
diffuse far-infrared emission.  There is an IRAS source,
IRAS 18462-0133, located within 20\arcsec\ of the cluster's
position.  The IRAS flux densities ranging from 16 Jy at 12 $\mu$m to
1500 Jy at 100 $\mu$m appear in Table~1, though they are highly uncertain 
due to the large beamsize in the high-background Galactic Plane.  

\subsection{Molecular Gas }

\begin{figure}
\vbox{
  \plotone{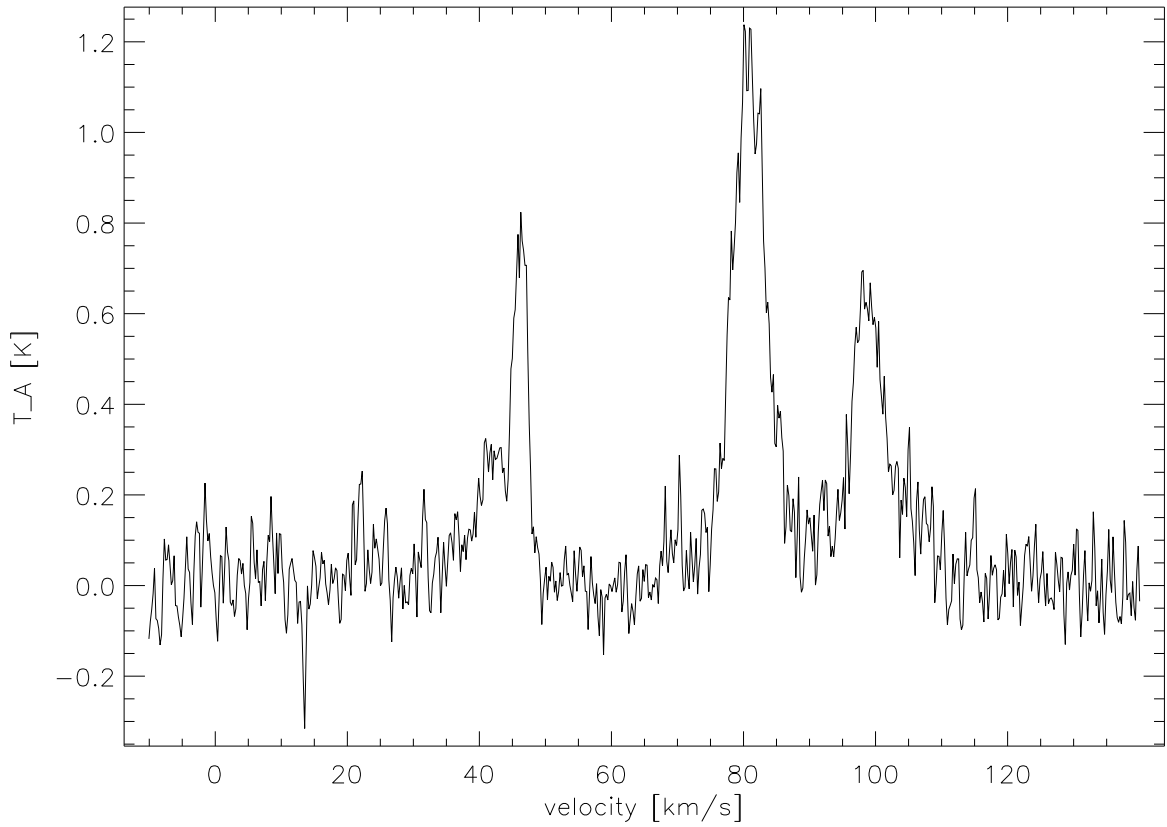}
} \figcaption[co] { \xiiico\ spectrum toward GLIMPSE-C01
from the Galactic Ring Survey (Simon \etal\ 2001).
Strong emission features near 46 \kms, 81 \kms, and 100 \kms\
correspond to near/far kinematic distances of 3.1/11.4 kpc, 5.2/9.3 kpc,
and 7.3 kpc, respectively.
  \label{co} }
\end{figure}

Millimeter-wave spectra from the \xiiico\ (1-0) Galactic Ring Survey
(GRS--Simon
\etal\ 2001; 46\arcsec\ beamsize) reveal strong emission at velocities
near 46 \kms, 81 \kms, and 100 \kms.  Figure~\ref{co} shows the
\xiiico\ spectrum toward the cluster.  The morphology of the molecular
feature near 46 \kms\ is similar to that seen in the 8.0 $\mu$m PAH
emission.  Figure~\ref{3colorCO} shows the IRAC 3.6 $\mu$m (blue), 5.8
$\mu$m (green), and 8.0 $\mu$m (red) images with the zeroth moment GRS
\xiiico\ map from 38 to 50 \kms\ in contours.  Contour levels denote
$I_{CO}=$3.0, 3.5, 4.0, 4.5, 5.0, 5.5, 6.5, 7.0, and 7.5 K
\kms.   Assuming a $^{12}CO/^{13}CO$ ratio of 40 (Langer \& Penzias
1990) and using $N_H{_2}(cm^{-2})= 2.0\times10^{20}~I_{^{12}CO}$
(Maloney \& Black 1988; Richardson \& Wolfendale 1988), these contours
correspond to molecular hydrogen columns of 2.5 --- 6.0$\times10^{22}$
\cmmb.  The corresponding extinction associated with this molecular
column density is $A_V=15$---$32$ for the molecular
component.\footnote{Here we assume $A_V=3.1\times E(B-V)= 3.1\times
N_{HI}/5.8\times10^{21}$ (Bohlin, Savage, \& Drake 1978). We convert
the $^{13}$CO column density to $^{12}$CO column density assuming a
ratio $^{12}$CO/$^{13}$CO=40 and then use
$N_H{_I}~(cm^{-2})=N_H{_2}~(cm^{-2})=2.0\times10^{22}~I_{^{12}CO}$.  } 
Note the CO peak which coincides with the diffuse IR clump 90\arcsec\
directly south (in Galactic coordinates) of the cluster.  This
morphological correspondence is one of the features which demonstrates
that the molecular emission is likely to be affiliated with the mid-IR
PAH emission.

\section{Analysis of the Cluster's Nature}

\subsection{Extinction}

Figure~\ref{CMD} shows a J-H versus H-K color-color
diagram of stars within 45\arcsec\ of the cluster center.  Lines
indicate the loci of the main sequence and giant branches for
$A_V=0$.  Dots are field stars from the 2MASS point
source catalog within an annulus between 1\arcmin\ and 9\arcmin\ from
the cluster.  Large symbols are the 225 stars within 45\arcsec\ of the
cluster measured in our WIRO JHK' photometry with photometric uncertainties
$<$0.1 mag in all three bandpasses.  The large star
designates the integrated photometry of the cluster. An arrow shows
the reddening vector for $A_V=15$ which is equivalent to $A_K=1.7$
using the extinction prescription of Cardelli, Clayton, \& Mathis
(1989).  The dotted box denotes the region occupied by probable cluster
members with similar colors.

Figure~\ref{CMD} shows that the vast majority of cluster stars are
consistent with 12-18 magnitudes of visual extinction,
similar to that estimated from the CO column density in the 46 \kms\ feature.
By comparison, field stars (dots) range from $A_V=0$ to $A_V>20$.  The range of
reddening among probable cluster members suggests patchy, variable
extinction.  We examined a reddening map of cluster stars and find no
large-scale extinction gradient across the face of the cluster.  Given
the presence of dust emission features in the 8.0 $\mu$m image near
the cluster, it is possible that dust mixed within the cluster
produces the variation in extinction among members.  

\subsection{Distance to the Cluster}

If the stellar cluster is located at the same distance as the mid-IR
PAH emission, then the correspondence with the CO indicates a
kinematic distance of either 3.1 kpc or 11.5 kpc (Clemens 1985).
However, even if the stellar cluster is not physically affiliated with the
PAH emission, we can still constrain distance to the cluster
using the observed extinction ($A_V\simeq15$ mag) and the kinematic
distances of the \xiiico\ features.  Given that the observed extinction
is similar to that implied by the \xiiico\ column density in the 46
\kms\ feature alone, the cluster must be at least 3.1 kpc away.  If the
cluster were more distant than the 81 \kms\ or 100 \kms\
\xiiico\ features, then the integrated \xiiico\ column densities 
would produce extinctions of $A_V>>30$ mag, in contradiction to the
observed $A_V\simeq15$ mag.  Thus, the cluster is on the {\it near}
side of the clouds producing the 81 \kms\ and 100 \kms\ emission.  The
kinematic distances of these features are 5.2/9.3 kpc, and 7.3
kpc (at the tangent point), respectively.  Therefore, 
the cluster must be closer than 7.3 kpc and 
{\it may} be
closer than 5.2 kpc if the 81 \kms\ \xiiico\ feature is located
at the near distance.  The total extinction along this sight-line 
estimated from COBE FIR maps is $A_V\simeq80$ mag corresponding to
$A_K=8.9$ (Schlegel \etal\ 1998).  This is somewhat lower than the
reddening estimate $A_V=200$ based on the \xiiico\ intensity of
$I_{^{13}CO}=23$ K \kms\ integrated over the entire Galactic velocity
range. In either case, the extinction estimated from Figure~\ref{CMD}
of $A_V\sim15$ is significantly lower than these maximum values,
suggesting that GLIMPSE-C01 is unlikely to be located on the far side
of the Galaxy and most likely lies in the range 3.1---5.2 kpc.

\subsection{Color Magnitude Diagram and K-band Luminosity Function}

Figure~\ref{CMD5} shows a K versus J-K color-magnitude diagram of
cluster stars. Lines illustrate the solar metallicity isochrones of
Bonatto, Bica, \& Girardi (2004) for ages of $10^8$ yr (solid), $10^9$
yr (dotted), and $10^{10}$ yr (dashed).  All isochrones have been
reddened by the equivalent of $A_V=15$.  Points are the stars within 45\arcsec of the
cluster which are probable cluster members based on color selection
criteria (those contained within the dotted box in Figure~\ref{CMD}.) 
The upper left, upper right, and lower left panels show the
theoretical isochrones at distances suggested by the kinematics of
\xiiico\ features: 3.1 kpc, 5.2 kpc, and 9.3 kpc.  At the two lower
distances, the tip of the $10^8$ yr isochrone lies well above the
magnitudes of the brightest cluster members and the mean color is not
well matched to the data.  Both the $10^9$ yr or $10^{10}$ yr
isochrones provide reasonable fits to the colors and magnitudes of
cluster stars at 3.1 and 5.2 kpc distances.  For this most probable distance
of 3.1 kpc, the data are
consistent with an old cluster in excess of 1 Gyr.  For the less
probable distance of 5.2 kpc in the upper right panel of
Figure~\ref{CMD5}, the $10^9$ yr or $10^{10}$ yr isochrones are still
in reasonable agreement with the data while the $10^{8}$ yr isochrone
is still a poor fit.  For the maximum possible distance of 9.3 kpc
(lower left panel) permitted by kinematic arguments, the tip of the
$10^{10}$ yr isochrone lies below the brightest cluster members, while
the $10^8$ yr and $10^9$ yr isochrones are now a reasonable fit both in
color and K-band magnitude. Given the dispersion in color among
cluster members, the strongest conclusion we can draw from
Figure~\ref{CMD5} is that either a nearby (3.1 kpc) old cluster or a
more distant young cluster may be consistent with the data.

Figure~\ref{lf} provides stronger constraints on the cluster age by
showing the K-band luminosity function of the cluster (thick line)
compared to expectations from the Bonatto \etal\ (2004) isochrones for
a cluster mass of $10^5$ \mo, a distance of 3.1 kpc, and three
different ages.  The K-band data have been corrected for 1.7 mag of
extinction ($A_V=15$).  The top panel shows luminosity functions for
$Z=0.019$ (approximately solar metallicity) clusters with ages of
$10^8$, $10^9$, and $10^{10}$ years.  The lower panel shows luminosity
functions for $Z=0.001$ (approximately 1/20 solar metallicity)
clusters with ages of $3\times10^9$, $6\times10^9$, and $10^{9}$
years.  The histogram of the GLIMPSE-C01 luminosity function has been
truncated at K=12.3 ($M_K=-2.3$) where the cluster photometry becomes
incomplete.  Figure~\ref{lf} shows that clusters of any metallicity
with ages $\leq3$ Gyr are inconsistent with the observed luminosity
function due to the lack of supergiants with luminosities $M_K<-6$ in
GLIMPSE-C01.  The best agreement between the models and data are for
clusters with ages $>3$ Gyr and masses from $1.0\times10^5$ to
$3\times10^5$ \mo, with the larger ages requiring larger masses.
Figure~\ref{lf} shows that a model with mass $3\times10^5$ \mo\ and
age $10^{10}$ yrs (lower panel, dashed line) provides an excellent fit
to the observed luminosity function.  Such masses and ages are
consistent with the canonical properties of globular clusters or
extraordinarily old, massive, open clusters.

\subsection{Luminosity}

Using the measured total fluxes in Table~1 along with the reddening of
$A_K=1.7$ mag and distance estimates derived above, we estimate a
total K-band luminosity $M_K=-10.3\pm0.6$ for the cluster at the near
3.1 kpc distance.  The corresponding V-band luminosity is $M_V=-8.4$,
assuming V-K=1.9 appropriate for a $10^9$ yr or $10^{10}$ yr
population (Leitherer \etal\ 1999).  GLIMPSE-C01 is, therefore, much
more luminous than Galactic open clusters (e.g., NGC~6791) and
approaches the luminosity of the most most massive globular clusters
(Harris 1996).  Adopting a greater distance would make GLIMPSE-C01
even more spectacular.  At the maximal distance of 9.3 kpc, it would
have $M_K=-12.8$ and $M_V=-10.9$ (assuming V-K=1.9), making it more
luminous than any known globular cluster.  Such luminosities are
typical of young (3-10 Myr) ``super star clusters'' found in nearby
starburst galaxies (reviewed by Whitmore 2000), but would be extraordinary 
for a globular cluster.

\subsection{Cluster Properties}

\begin{figure}
\vbox{
  \plotone{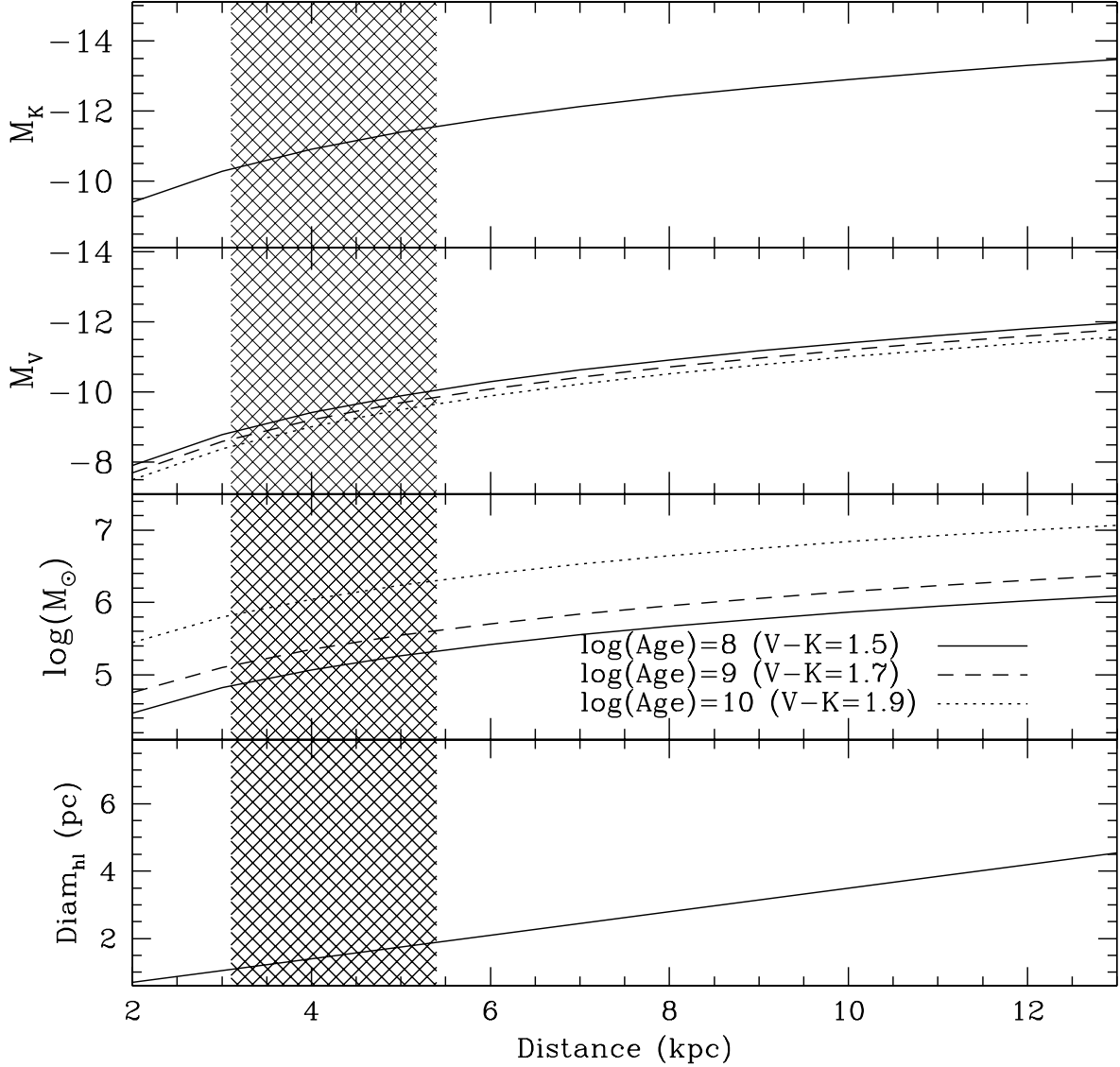}
} \figcaption[dist] {Cluster properties versus assumed distance.
The most probable distances between 3.1 and 5.2 kpc is shaded.
The V-band luminosity ranges between $M_V=-8.4$ and $M_V=-10.0$ for
plausible distances.
The total mass is $10^5$ \mo\ to $10^{6}$ \mo\
for the most probable ages and distances.  \label{dist} }
\end{figure}

Figure~\ref{dist} illustrates how the derived physical properties of
GLIMPSE-SC01 scale with distance.  The shaded region highlights the
most probable distance.  The upper two panels show that the inferred
K-band and V-band luminosities are $M_K=-10.3$ and $M_V=-8.4$ for
distances near 3.1 kpc.  For the larger but less probable distances,
these luminosities would be correspondingly larger.  Here, the V-band
luminosities have been derived from the measured K-band luminosities
using the theoretical colors, V-K=1.9, for a instantaneous burst
stellar population with age $10^9$---$10^{10}$ yrs (Leitherer \etal\
1999).  The mass of the cluster in the third panel is inferred from
the theoretical mass to light ratio (tabulated by Bonatto, Bica \&
Girardi 2004) for three different representative ages of $10^8$,
$10^9$, and $10^{10}$ yrs.  In any case, the mass of the cluster
exceeds $10^5$ \mo\ for all reasonable distances and surpasses $10^6$
\mo\ if the age is very old or the distance is large.  The
conservative mass estimate of $10^5$ \mo\ is similar to Galactic
globular clusters and young super star clusters with measured
dynamical masses in nearby starbursts (e.g., NGC~1569--Ho \&
Filippenko 1996).  The lower panel of Figure~\ref{dist} indicates that
the half light diameter of the cluster falls in the range 1-2 pc.
This is smaller than most globular clusters
which have half light diameters of 3-6 pc (Harris 1996).  
It is possible that the passage of GLIMPSE-C01 through the 
Galactic disk, or a prolonged presence near the disk, has
stripped some fraction of the outer, loosely bound cluster members
and left only the tightly bound cluster core.   

\section{Discussion of Why GLIMPSE-C01 is Likely to be a Globular Cluster}

Several lines of evidence favor identifying GLIMPSE-C01
as a member of the classical Milky Way globular cluster system rather than a
young stellar cluster or even an evolved old open cluster.

\begin{figure}
\vbox{
  \plotone{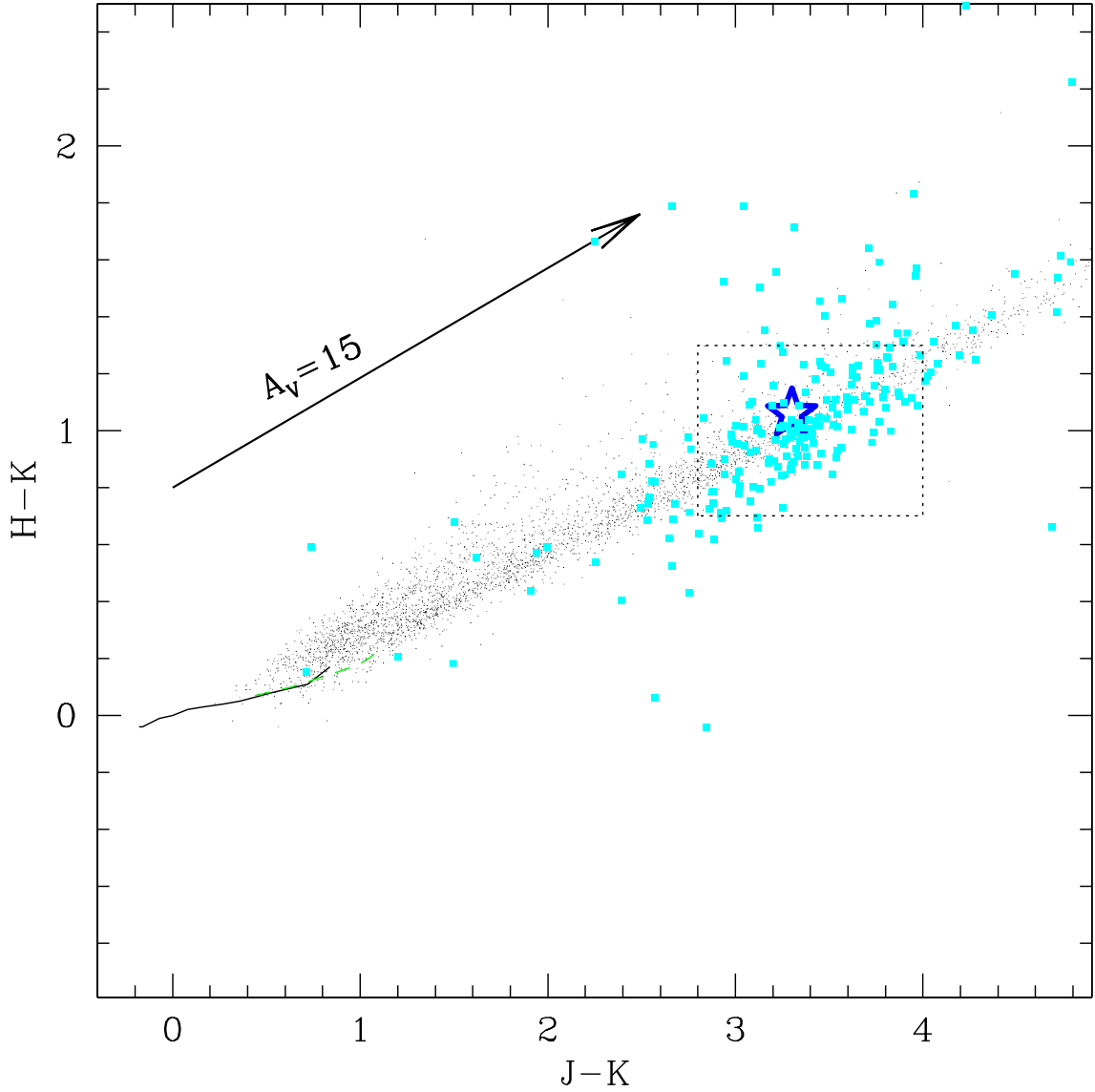}
} \figcaption[CMD] {JHK' color-color diagram 
of stars within 45\arcsec\ of the cluster center.  Lines show the
loci of main sequence and giant stars.   Dots are
field stars from the 2MASS catalog in an annulus between 
1\arcmin\ and 9\arcmin\
from the cluster.   Large points are stars within the 45\arcsec\ of the
cluster center as measure with WIRO/GIRcam photometry.  The large star  
designates the integrated colors of the cluster.
An arrow displays the reddening vector for $A_V=15$.  The dashed box
encloses probable cluster members selected on the basis of
similar reddening.  
  \label{CMD} }
\end{figure}

\begin{figure}
\vbox{
  \plotone{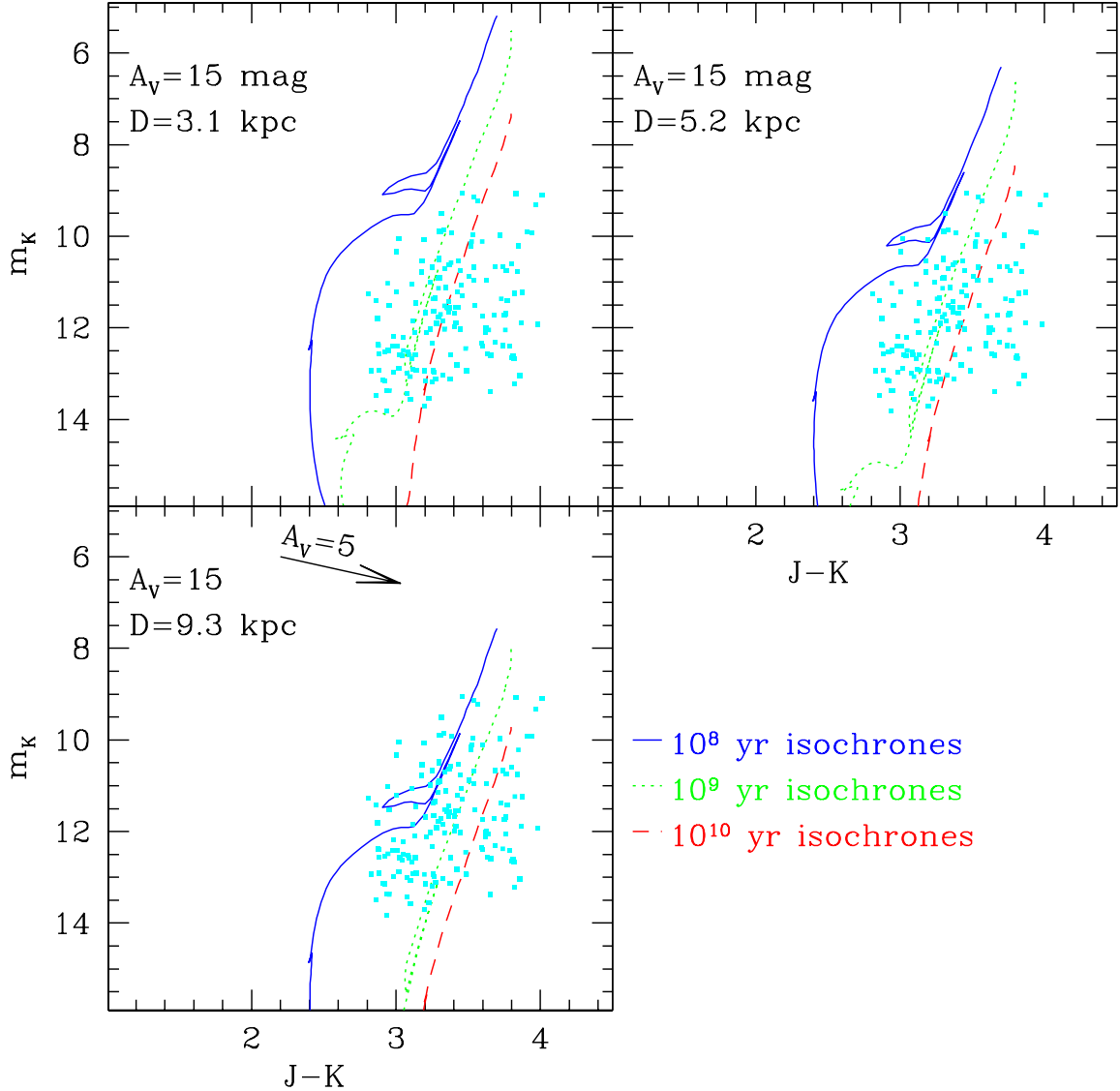}
} \figcaption[CMD5] {K versus J-K color-magnitude diagram 
of stars within 45\arcsec\ of the cluster center
from within the dotted box in Figure~\ref{CMD}.  Lines show the
theoretical isochrones of Bonatto, Bica, \& Girardi (2004)
for ages of $10^8$ yrs (solid), $10^9$ yrs (dotted), and $10^{10}$ yrs
(dashed).   The three panels compare the isochrones to WIRO photometry
for distances of 3.1, 5.2, and 9.3 kpc.  The best fit is achieved
with the  $10^9$ yr or $10^{10}$ yr isochrones at a distance of 3.1 -- 5.2 kpc,
consistent with the kinematic \xiiico\ distance.
  \label{CMD5} }
\end{figure}

\begin{figure}
\vbox{
  \plotone{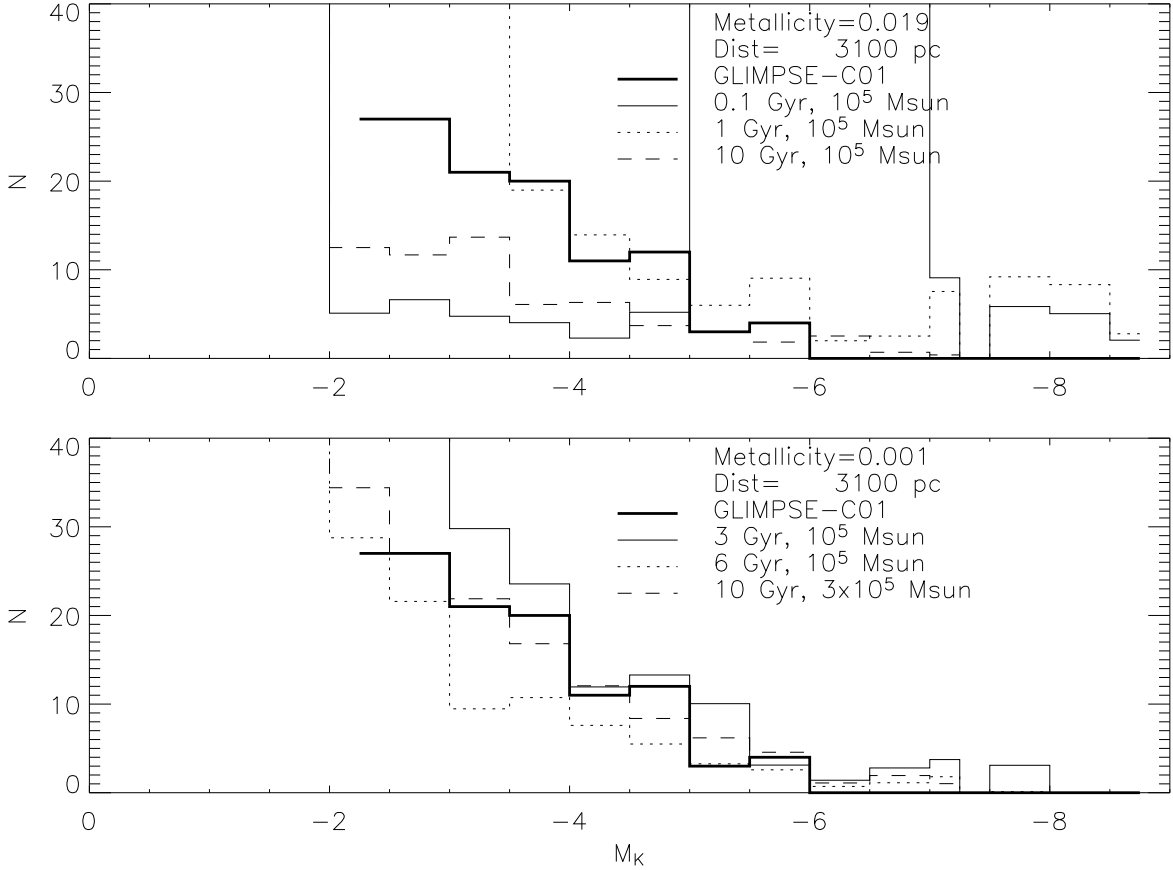}
} \figcaption[lf] {K-band luminosity function of GLIMPSE-C01 (thick line)
compared to theoretical luminosity functions from the
Bonatto, Bica, \& Girardi
(2004) isochrones for a cluster mass of $10^5$ \mo,
a distance of 3.1 kpc, and three different ages.  The K-band data
have been corrected for 1.7 mag of extinction ($A_V=15$).  
The top panel shows luminosity functions for $Z=0.019$ 
(approximately solar metallicity)
clusters with ages of $10^8$, $10^9$, and $10^{10}$ years.
The lower panel  shows luminosity functions for $Z=0.001$ 
(approximately 1/20 solar metallicity)
clusters with ages of $3\times10^9$, $6\times10^9$, and $10^{9}$ years.
Clusters of any metallicity with ages $\leq3$ Gyr are inconsistent
with the observed luminosity function due to the lack of supergiants
with luminosities $M_K<-6$ in GLIMPSE-C01.
 \label{lf} }
\end{figure}

\begin{figure}
\vbox{
  \plotone{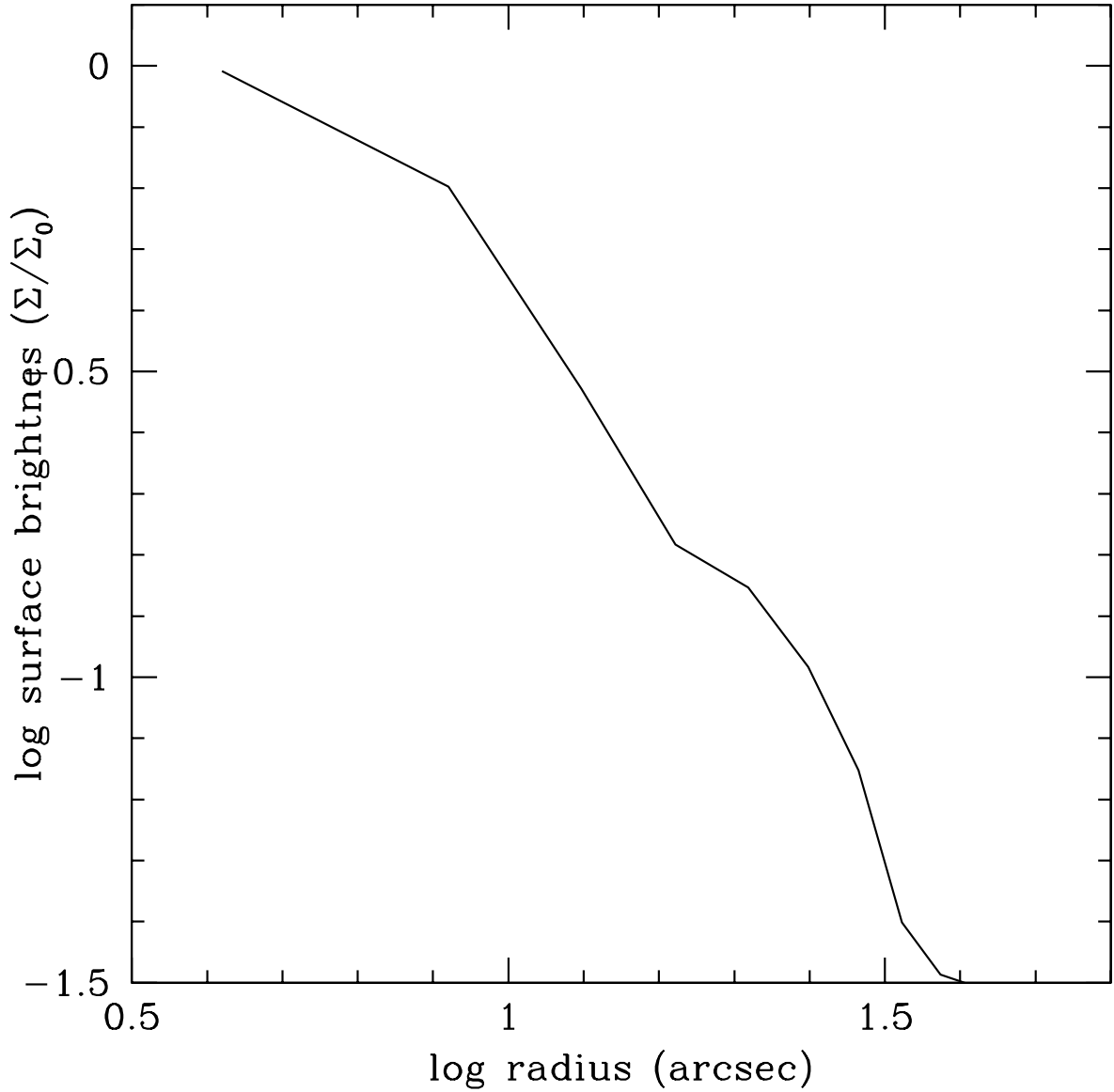}
} \figcaption[profile] {The
3.6 $\mu$m surface brightness as a function of radius for GLIMPSE-C01,
normalized to the central surface brightness.  At large radii the
surface brightness becomes very uncertain due to contamination by
field stars in the Galactic Plane.
  \label{profile} }
\end{figure}

\begin{itemize}
\item{Lack of Radio Emission: Prominent Galactic star formation
regions are luminous thermal radio and infrared sources with fluxes of
many tens or even hundreds of Janskys (e.g., Westerlund 2/RCW49--
Churchwell \etal\ 2004). A modest star forming region like Orion has
an integrated 1420 GHz flux of 420 Jy at a distance of 450 pc (Felli
\etal\ 1993).  Orion would have a radio continuum flux of 9.4 Jy at 3.1
kpc, the adopted distance of GLIMPSE-C01, or a flux of 225 mJy if it
were on the far rim of the Galaxy.  The new cluster, by comparison, is
a marginal detection at $<$5 mJy in the radio continuum, 
indicating that no massive stars are present.  The
lack of radio synchrotron sources or diffuse infrared shells suggests
that we can rule out recent supernova remnants.  }
\item{A well-populated giant branch:  Given the infrared 
photometry presented in 
Figure~\ref{CMD} and the K-band luminosity function in
Figure~\ref{lf}, the cluster contains a wealth of giant stars
but a lack of luminous supergiants,
consistent with the evolved nature of a globular cluster.  
Preliminary K-band spectroscopy of GLIMPSE-C01 shows no
emission lines which should be present in a very young OB type cluster 
(D. Clemens, in preparation).}
\item{Overall luminosity:  With an absolute V magnitude estimated 
at $M_V=-8.4$,
the cluster is more luminous than the majority of known globular
clusters (Harris 1996) and vastly more luminous and massive
than known old open clusters even at the conservative distance of 3.1 kpc.}
\item{Stellar density:  GLIMPSE-C01 is a rich, centrally condensed
cluster with evidence for mass segregation characteristic of
dynamically relaxed systems.  Figure~\ref{profile} shows the
3.6 $\mu$m surface brightness as a function of radius,
normalized to the central surface brightness.  At large radii the
surface brightness becomes very uncertain due to contamination by
field stars.
 Open clusters, in general,
do not survive long enough to become dynamically relaxed.
Figure~\ref{compare} compares the IRAC 4.5 $\mu$m image of 
GLIMPSE-C01 (left) with the 2MASS K-band image of the old
open cluster NGC~6791 (right) which lies at a similar distance (4 kpc).}
NGC6791 is $\sim7$ Gyr old (Demarque, Guenther, \& Green 1992) 
but not as rich or centrally condensed as the new cluster, even though
it is at a similar distance (4.0 kpc).
\item{Position in the Galaxy:  At $leq$6.1 kpc from
the Galactic center, GLIMPSE-C01 is interior to known old ($>$1 Gyr)
open clusters
which are only found at $R_{GC}> 7.5$ kpc (Friel 1995).  Survival
over Gyr timescales requires that a weakly bound old open cluster
be protected from disruption by the gravitational 
forces of giant molecular clouds and stellar encounters which are
more prevalent inside the solar circle.   Globular clusters,
on the other hand, are preferentially found toward the Galactic bulge.}
\end{itemize}

\noindent Given the evidence, it appears likely that GLIMPSE-C01 is a massive
globular cluster making a passage through the Galactic disk.  The
morphology of the cluster and surrounding ISM, if physically
associated with the cluster, may provide some clues regarding its
trajectory.  

The outer isophotes of GLIMPSE-C01 are significantly elliptical,
having $e=(1-b/a)=0.21$ at 72\arcsec\ from the nucleus.
Figure~\ref{ellipse} shows the IRAC 3.6 $\mu$m image with a series of
best fit ellipses having semi-major axes of 12, 24, 36, 48, 60, and 72
arcseconds.  The ellipticity of the outer ellipses shows that the
major axis of the cluster lies along position angle 124\degr\ in Galactic
coordinates or 61\degr\ in J2000 equatorial coordinates.

The bright ``plume'' of PAH emission seen in Figure~\ref{3color} lies
at the same position angle as the major axis of the cluster's outer
isophotes, suggesting a preferred orientation at a position angle of
$\sim124$\degr\ from Galactic north (PA=61\degr\ in equatorial
coordinates).  One end of the plume is centered on the cluster and the
other end extends 1.5\arcmin\ to the Galactic south of the cluster and
toward higher longitudes.  One possibility is that the plume traces
intracluster or circumstellar debris from the GLIMPSE-C01 which has
been stripped by the Galactic ISM as the cluster moves from south to
north across the the Plane.  An alternative scenario is that the
diffuse ``ribbon'' of emission extending to the Galactic north and
toward lower longitudes traces the recent trajectory of the cluster
and that the plume is caused by recent star formation as the cluster
impacts the Galactic molecular layer on its journey to the south.  The
possibility remains, of course, that the diffuse IR emission lies in
the foreground or background and is not affiliated with the cluster.

\begin{figure}
\vbox{
  \plotone{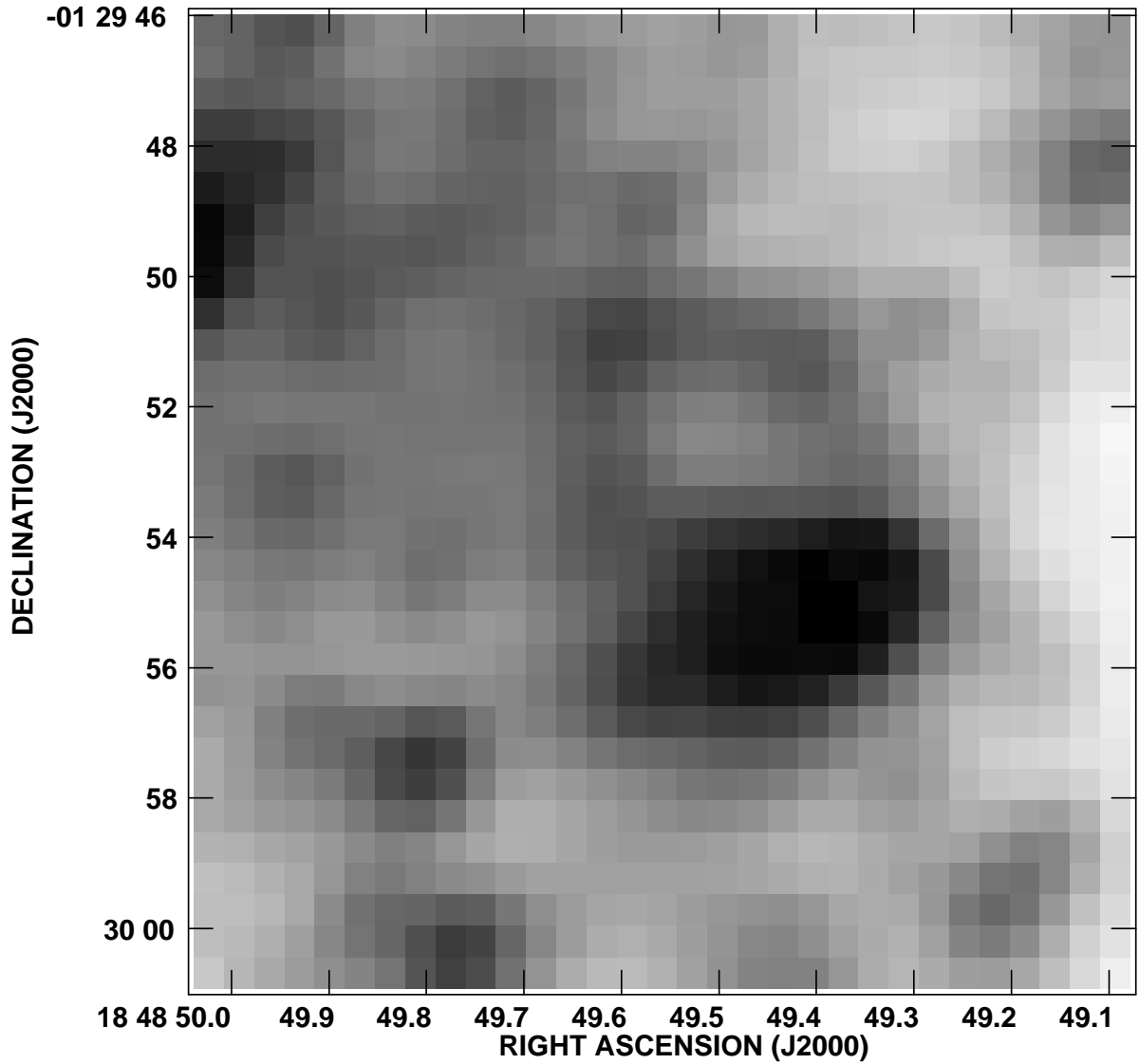}
} \figcaption[wiro] {A subsection of the WIRO H-band image of
the GLIMPSE-C01 showing a loop-like structure which is seen at all 
near- and mid-infrared continuum bands.    The loop has
a diameter of about 9 pixels, corresponding to
a linear diameter of 
0.058 pc or $\sim12,000$ AU at 3.1 kpc.  This object may be an
old nova shell or a young planetary nebula.   
  \label{wiro} }
\end{figure}

One interesting object visible in both the WIRO near-IR and IRAC
mid-IR images is a loop-like structure on the Galactic north side of the
brightest stellar peak in the cluster.  The center of the loop is
located at RA(2000)=18h48m49.58s, DEC(2000)=-01d29m53s.  
Figure~\ref{wiro} shows the WIRO H-band image of a small
region around this feature.  The loop has
a diameter of about 9 pixels (4) arcsec in the WIRO images.  It is
seen at all wavelengths except 8$\mu$m where the angular resolution of
\spitz/IRAC becomes insufficient to resolve it.  At the adopted
distance of the cluster, the linear diameter of the loop would be
0.058 pc or $\sim12,000$ AU.  This size is much larger than dust
shells ejected by individual asymptotic giant branch stars near the
end of their lives.  The lack of strong radio emission rules out a 
supernova remnant.  The shell diameter is comparable to the
dimensions of old nova shells with ages of tens of years (e.g., Downes
\& Duerbeck 2000) which are generally seen in emission lines, but
sometimes detected in continuum light.  Dust in the shell could
scatter stellar light from the cluster and explain the strong
continuum detection.  The object may also be a young planetary nebula
shell.  In either case, velocity-resolved spectroscopy should be
performed to measure the kinematics of the shell and provide
additional clues to its origin.  If the object is a recent nova
shell produced by a cluster member, and if archival images exist that
could pinpoint the date of the nova, then a measured expansion
velocity could be used to measure a direct kinematic distance to the
cluster independent of assumptions about a Galactic
rotation curve or peculiar cluster motions.  However, with
optical extinctions of $A_V=15$, even the brightest Galactic 
novae with absolute visual magnitudes of -11 (Duerbeck 1981)
would only reach apparent magnitudes of $V=16$, making it unlikely
that it would have been observed. 
If the object is a planetary nebula, chemical analysis of
the nebula would help to determine whether GLIMPSE-C01 belongs to the
older, metal-poor halo population of globular clusters or the younger
more metal-rich disk population.

This discovery of a new nearby globular cluster suggests a plethora of
followup observational programs.  High resolution infrared
spectroscopy of cluster members would be highly desirable to estimate
its radial velocity and establish a kinematic distance.  A proper
motion measurement is needed in order to compute its orbital
parameters and determine a probable
origin in either the old halo cluster population of the younger disk
population.  Deep wide-field infrared photometry would help to
establish the boundaries of the cluster and better constrain its total
luminosity.  The region of Y-shaped stellar outflows to the south (Galactic)
of the cluster will make an interesting target for multi-wavelength
studies of star formation activity in the vicinity of the cluster.

\acknowledgments

We thank the anonymous referee for a timely and helpful review.
We are grateful to Stephan Jansen for his invaluable work maintaining
the GLIMPSE computing network.  We thank Jim Weger and Phil Haynes for
assistance at WIRO.  John Stauffer contributed key insights on an early
version of this manuscript.  We thank Ata Sarajedini for suggesting
analysis of the K-band luminosity function which led to a more robust
distance and age estimate.  Support for this work, part of the
\spitz\ Space Telescope Legacy Science Program, was provided by NASA
through Contract Numbers (institutions) 1224653 (UW), 1225025 (BU),
1224681 (UMd), 1224988 (SSI), 1259516 (UCB), 1253153 (UMn), 1253604
(UWy), 1256801 (UWW) by the Jet Propulsion Laboratory, California
Institute of Technology under NASA contract 1407. Brian Uzpen was
supported by a Wyoming NASA Space Grant Consortium, NASA Grant
NGT-40102 40102, Wyoming NASA EPSCoR Grant NCC5-578 and 1253604. This
publication makes use of data products from the Two Micron All Sky
Survey, which is a joint project of the University of Massachusetts
and the Infrared Processing and Analysis Center/California Institute
of Technology, funded by the National Aeronautics and Space
Administration and the National Science Foundation.
The Two Micron All Sky Survey is a joint
project of the University of Massachusetts and the Infrared Processing
and Analysis Center/California Institute of Technology, funded by the
National Aeronautics and Space Administration and the National Science
Foundation."

\clearpage

\begin{figure}
\vbox{
} \figcaption[IRAC] {Logarithmic greyscale images of the cluster at each 
of the four IRAC bandpasses: 
3.6 microns (upper left), 4.5 microns (upper right), 5.8 microns (lower left),
and 8.0 microns (lower right).  The IRAC 1 \& 2 bands (upper
panels) are dominated by stellar photospheric emission.   Stars become less
prominent in IRAC bands 3 \& 4 which are are increasingly dominated by
emission from known polycyclic aromatic hydrocarbon (PAH) bands.
  \label{IRAC} }
\end{figure}

\clearpage

\begin{figure}
    \caption{Three-color mid-infrared image of the cluster constructed from
the IRAC 8.0 micron image (red), IRAC 5.8 micron image (green) and the IRAC 3.6 micron 
image (blue).  Contours show the 1420 GHz radio continuum emission from
the NRAO VLA Sky Survey at multiples of 2,3,4,5, and 10 times the
1$\sigma$ RMS sensitivity of 1.3 mJy/beam.  The marginal radio
detection coincident with the cluster has a peak flux density of 
4.5 mJy/beam.   \label{3color} }
\end{figure}

\clearpage

\clearpage

\clearpage

\begin{figure}
    \caption{Three-color mid-infrared image of the cluster constructed from
the IRAC 8.0 micron image (red), IRAC 5.8 micron image (green) and the IRAC 3.6 micron 
image (blue).  Contours show the \xiiico\ (1-0) molecular
line emission from the Galactic Ring Survey 
(45\arcsec\ beamsize) integrated over the LSR 
velocity range 37-50 \kms.  
Contours denote levels of $I_{^{13}CO}=$
3.0, 3.5, 4.0, 4.5, 5.0, 5.5, 6.5, 7.0, and 7.5 K \kms\ which correspond
to molecular hydrogen columns of 2.5$\times10^{22}$--- 
6.0$\times10^{22}$ \cmmb\
(see text for details).
The correspondence between diffuse PAH emission seen at 8.0 $\mu$m
and \xiiico\ suggests a common origin and motivates adoption of the kinematic
distance of 3.1 kpc.   
  \label{3colorCO} }
\end{figure}

\clearpage

\clearpage

\clearpage

\clearpage

\clearpage

\clearpage

\begin{figure}
\vbox{
} \figcaption[compare] {A comparison of the richness of
the putative globular cluster GLIMPSE-C01 seen at 4.5 $\mu$m 
to the old open cluster NGC~6791
(D=4 kpc) from the 2MASS K' image (right).
  \label{compare} }
\end{figure}

\clearpage

\begin{figure}
\vbox{
} \figcaption[ellipse] {IRAC 3.6 $\mu$m image of GLIMPSE-C01 with a series of
best fit ellipses having semi-major axes of 12, 24, 36, 48, 60, and 72
arcseconds.   The line labeled `N' designates north in
equatorial coordinates.   The outer two ellipses show that the
cluster is elongated with ellipticity $e=1-(b/a)=0.2$ 
at position angle 124\degr\ in Galactic coordinates
(61\degr\ in J2000 equatorial coordinates).   
  \label{ellipse} }
\end{figure}

\clearpage

\clearpage

\begin{deluxetable}{lc}
\tablecaption{Cluster Parameters}
 \label{data.tab}
\tablewidth{6.5in}
\setlength{\tabcolsep}{0.12in} 
\tablehead{
\colhead{Parameter} & 
\colhead{Value} }
\startdata
RA (2000)     & 18h48m49.7s     \\
DEC (2000)    &-01d29m50s     \\
$l$           & 31.30    \\
$b$           & -0.10    \\
$R_\odot$ (kpc)\tablenotemark{a}     & 3.1$\pm0.5$    \\
$R_{GC}$ (kpc)\tablenotemark{a}      & 6.8    \\
$r_h$ (arcsec)\tablenotemark{b}      & 36  \\
$r_c$ (arcsec)\tablenotemark{c}      & 30  \\
$m_K$ (mag)\tablenotemark{d}               & 3.77 (19.3 Jy)  \\
$J-H$ (mag)\tablenotemark{d}               & 2.24 \\
$J-K$ (mag)\tablenotemark{d}               & 3.30    \\
$F_{3.6} (Jy)$\tablenotemark{d}           & 14.3   \\
$F_{4.5} (Jy)$\tablenotemark{d}           & 9.9   \\
$F_{5.8} (Jy)$\tablenotemark{d,e}           & 14.6  \\
$F_{8.0} (Jy)$\tablenotemark{d,e}           & 23.7   \\
$F_{12} (Jy)$\tablenotemark{f}           & 16.4L   \\
$F_{25} (Jy)$\tablenotemark{f}           & 18.9:  \\
$F_{60} (Jy)$\tablenotemark{f}           & 285   \\
$F_{100} (Jy)$\tablenotemark{f}           & 1516L   \\
$A_V$ (mag)                          & 15$\pm3$   \\
$A_K$ (mag)                          & 1.7$\pm0.3$   \\
$m_{K_0}$\tablenotemark{g}           & 2.07$\pm0.3$    \\
$V-K$\tablenotemark{h}               & 1.5 -- 1.9   \\
$M_K$\tablenotemark{i}               & -10.3$\pm0.6$    \\
$M_V$\tablenotemark{i}               & -8.4$\pm3$   \\

\enddata
\tablenotetext{a}{estimated distance from the sun and distance from the Galactic center}
\tablenotetext{b}{half light radius at K, 3.6 $\mu$m, 4.5 $\mu$m}
\tablenotetext{c}{core radius, defined as the radius where the surface brightness
	drops to half of the central value }
\tablenotetext{d}{apparent magnitude or flux, not corrected for extinction, within 
	a 90 \arcsec\ radius aperture}
\tablenotetext{e}{Note: the flux in IRAC bands 3 \& 4 contains a considerable
	contribution from diffuse PAH emission which is not present at other wavelengths}
\tablenotetext{f}{IRAS flux; most measurements are highly uncertain due to 
large beamsize and high background}
\tablenotetext{g}{apparent magnitude, corrected for extinction, with $A_K=1.7$ mag}
\tablenotetext{h}{estimated V-K color for starbursts of ages $10^8$ yrs
to $10^{9}$ yrs (Starburst99) and for globular clusters (Harris 1996)}
\tablenotetext{i}{estimated absolute magnitudes at a distance of 3.1 kpc,
including uncertainties on distance and reddening}
\tablenotetext{j}{probable apparent integrated
V magnitude given the adopted extinction and V-K color}
\end{deluxetable}

\end{document}